\documentclass[multphys,vecphys]{svmult}

\usepackage{makeidx}         
\usepackage{graphicx}        
\usepackage{multicol}        
\usepackage[bottom]{footmisc}

\makeindex             

\begin{document}

\title*{Infrared Surface Brightness Analysis of Galaxies in Compact Groups}
\author{Ilse Plauchu-Frayn, Roger Coziol \and
H\'ector Bravo-Alfaro.}
\institute{Departamento de Astronom\'ia, Universidad de Guanajuato, Apdo. Postal 144, 36000 Guanajuato, M\'exico 
\texttt{plauchuf@astro.ugto.mx},
\texttt{rcoziol@astro.ugto.mx},
\texttt{hector@astro.ugto.mx}}
%
%
\maketitle

Observations were carried out during seven nights at the 2.12m telescope of the Observatorio Astron\'omico Nacional, located in Baja Califormia, M\'exico. The images were obtained with the NIR Camera CAMILA using J and K' filters for seven groups (Plauchu-Frayn et al. 2006). 


Our analysis follows the method described in Barth et al. (1995). We used the program ELLIPSE in IRAF to fit ellipses to NIR isophotes. The program yields the ellipticity, position angle and variation of ellipticity through the higher-order harmonics a4 and b4. Strong deviations from ellipticity and of position angle indicate inhomogeneous mass distributions related to tidal interactions (plumes and tail), mergers, bars or presence of a companion. 

  In general, the level of perturbation (a4) is well correlated with the level of activity. The trends observed seem to follow the spectroscopic classification found by Coziol et al. (2004): the level of perturbation decreases from A to B to C. The CGs of types A and B look much less "evolved" than those of type C, which suggests they formed more recently. 
  
     The lack of perturbation in HCG 74 may suggest the galaxies are located in deeper potential wells, which would be consistent with the high velocity dispersion of the group. This is as expected if they are the results of past mergers, also consistent with the earlier morphological types and absence of gas. However, our analysis does not exclude the possibility of an ongoing merger phase, only that interactions need more time to produce visible perturbations. Alternatively, massive systems are expected to form first in more massive dark matter haloes, and the galaxies could thus be in equilibrium. More observations of CGs of this type will be necessary to distinguish between the two possibilities.

%
\begin{figure}
\centering
\includegraphics[height=18cm]{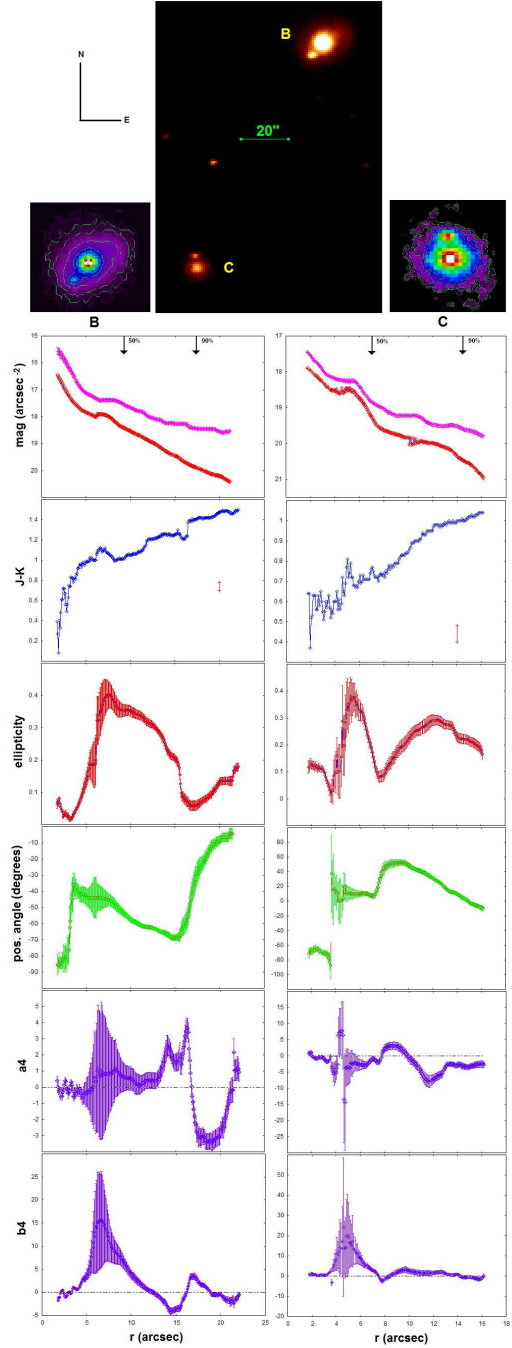}
%
%
\caption{One example of our analysis: members b y c of HCG88 (type A) in J band and two of its members (b and c) in false colours and isophotes. The green line indicates the image scale. From top to bottom we plot the surface brightness, colour index and geometrical parameters: ellipticity, position angle and harmonics a4 y b4, as a function of the semimajor axis length r. The 50\% (R50) and 90\% (R90) isophotal radii are also indicated. Both b and c show signs of merger with a small companion within R90 and R50, respectively. In the c member, the high level of perturbations is correlated with a strong starburst activity. Both members also present strong perturbations beyond R90 due to star formation in the disks.}

\label{fig:1}       
\end{figure}
%
%
%
%
%

%
%



\printindex
\end{document}